\documentclass[graybox]{svmult}

\usepackage{type1cm}        % activate if the above 3 fonts are
\usepackage{makeidx}         % allows index generation
\usepackage{graphicx}        % standard LaTeX graphics tool
\usepackage{multicol}        % used for the two-column index
\usepackage[bottom]{footmisc}% places footnotes at page bottom

\usepackage{newtxtext}       % 
\usepackage[varvw]{newtxmath}       % selects Times Roman as basic font

\usepackage{hyperref} % Para poner links
\usepackage{multirow}
\usepackage{color}
\usepackage{colortbl}

\makeindex             % used for the subject index
\begin{document}

\title*{Machine learning applied to omics data}
\author{Aida Calviño, Almudena Moreno-Ribera and Silvia Pineda}
\institute{Aida Calviño \at Department of Statistics and Data Science, Complutense University of Madrid, Madrid, Spain \email{aida.calvino@ucm.es}
\and Almudena Moreno-Ribera \at Department of Statistics and Data Science, Complutense University of Madrid, Madrid, Spain \email{almmor02@ucm.es}
\and Silvia Pineda \at Department of Statistics and Data Science, Complutense University of Madrid, Madrid, Spain \email{sipineda@ucm.es}}
\maketitle

\abstract{In this chapter we illustrate the use of some Machine Learning techniques in the context of omics data. More precisely, we review and evaluate the use of Random Forest and Penalized Multinomial Logistic Regression for integrative analysis of genomics and immunomics in pancreatic cancer. Furthermore, we propose the use of association rules with predictive purposes to overcome the low predictive power of the previously mentioned models. Finally, we apply the reviewed methods to a real data set from TCGA made of 107 tumoral pancreatic samples and 117,486 germline SNPs, showing the good performance of the proposed methods to predict the immunological infiltration in pancreatic cancer.}

% \abstract{Each chapter should be preceded by an abstract (no more than 200 words) that summarizes the content. The abstract will appear \textit{online} at \url{www.SpringerLink.com} and be available with unrestricted access. This allows unregistered users to read the abstract as a teaser for the complete chapter.\newline\indent
% Please use the 'starred' version of the \texttt{abstract} command for typesetting the text of the online abstracts (cf. source file of this chapter template \texttt{abstract}) and include them with the source files of your manuscript. Use the plain \texttt{abstract} command if the abstract is also to appear in the printed version of the book.}

\section{Introduction}
\label{sec:1}

Big data has revolutionized the biomedical field and the way we study complex traits resulting in the generation of vast amounts of omics data, such as genomics, epigenomics, transcriptomics, microbiomics or immunomics among others. 

In this era of rapid molecular technological advancements, studies using classical statistical techniques are becoming too simplistic when considering the reality of complex traits. Instead, Machine Learning (ML) techniques that may reflect combinatorial effects (including additive and interactive) should be contemplated to address such complexity. Given the wealth and availability of omics data, ML methods provide a powerful opportunity to improve the current knowledge about the determinants of complex diseases \cite{reel2021machine}. Developing efficient strategies to identify new markers and its interactions requires applying ML techniques through the integration of large omics data \cite{ritchie2015methods,lopez2019challenges,pineda2020data}. 

Many recent findings have been powered by statistical advances that have enabled profiling
of the biological system on multiple levels. Through gene expression, genetics, immune repertoire and cell profiling, we can characterize potential new markers playing decisive roles in health and disease. In this regard, the immune system requires special attention. We now know that the immune system is not just responsible for protection against diseases and host defense but also plays a role in tissue maintenance and repair. As such, the system is spread throughout not only the blood and lymphatic systems but also most tissues. The immune system plays a key role in the regulation of tumor development and progression, and crosstalk between cancer cells and immune cells has been incorporated into the list of major hallmarks of cancer \cite{sehouli2011epigenetic}, consequently, the tumor microenvironment is closely connected to every step of tumorigenesis \cite{wang2017role}. Translating this complexity into new biological insights requires advanced statistics, data mining, and ML skills. 

In this chapter, we focus on the proper use of ML techniques applied to the study of omics data in pancreatic cancer (PC). PC is a dreadful disease and it is the deadliest cancer worldwide with 7\% 5-year survival rate. Despite the efforts to advance with this disease in many angles, there are still a number of unknown risk factors and its interactions and correlations contributing to this devastating disease. Immunological infiltration and its interactions with other factors plays an important role in PC development \cite{barnes2017hype}. Indeed, some years ago, it was shown that genetic variations explained an appreciable fraction of trait heritability \cite{orru2013genetic}, and very recently it has been estimated that environmental factors and genetic susceptibility may explain 50\% and 40\% of the immune system differences \cite{patin2018natural,liston2018origins} respectively, suggesting the feasibility of genetic dissection of quantitative variation of specific immune cell types. Thus, we can hypothesize that the characteristics of the tumoral immune infiltration are modulated by genetic susceptibility.

This chapter focuses on the use of ML techniques to integrate genomic-immunomic data that may be useful in deciphering new biological insight that will help to advance PC research. First, we describe the omics data used in this work in Section \ref{sec:2} and discuss the main challenges faced by these data types in Section \ref{sec:3}. Section \ref{sec:4} is devoted to the most important ML techniques in the biomedical field, as well as others less known. Finally, we show an application to a real data set in Section \ref{sec:5} and give some conclusions and lines of future work in Section \ref{sec:6}.

\section{Data types}
\label{sec:2}

\subsection{Genomics}
\label{subsec:2.1}

The whole genetic information of an individual can be studied using whole genome sequencing determining the order of all nucleotides within the DNA molecule. Most of the studies have determined a subset of genetic markers to capture as much of the complete information as possible using genotyping information. These markers are Single Nucleotide Polymorphisms (SNPs) that are changes of one nucleotide base pair that occur in at least 1\% of the population. In humans, the majority of the SNPs are bi-allelic, indicating the two possible bases at the corresponding position within a gene. If we define A as the common allele and B as the variant allele, three combinations are possible: AA (the common homozygous), AB (the heterozygous) and BB (the variant homozygous) which can be transformed into $0,1,2$ considering the number of variant alleles. These combinations are known as the genotypes and they are assessed using either genotyping platforms or  high throughput sequencing generally from germline DNA (see Figure \ref{fig:1}A). One very well known type of study that utilizes genotyping information are the Genome-wide association studies (GWAS) used to identify genetic variants (SNPs) extracted from germline DNA that are significantly associated with disease states (healthy vs. unhealthy individuals). 

Using GWAS in combination with ML techniques, advances in the identification of  genetic susceptibility loci have been shown in cancer \cite{behravan2018machine}, sclerosis \cite{zhang2022genome-wide} or covid-19 \cite{downes2021identification} and many other diseases. How such variation might relate to a particular disorder is not always known and more complex analyses are needed. Linking particular genetic variants (SNPs) to associated mechanisms, and one directly related with the omics integration is through the expression of quantitative trait loci (eQTL) analysis, which are SNPs that partly explain the variation of a gene expression phenotype. In recent years, many statistical and ML methods for eQTL analysis have been developed with the ability to provide a more complex perspective towards the identification of relationships between genetic variation and genetic expression \cite{chen2019statistical}.

\subsection{Immunomics}
\label{subsec:2.2}

\begin{figure}[t]
\includegraphics[width=\textwidth]{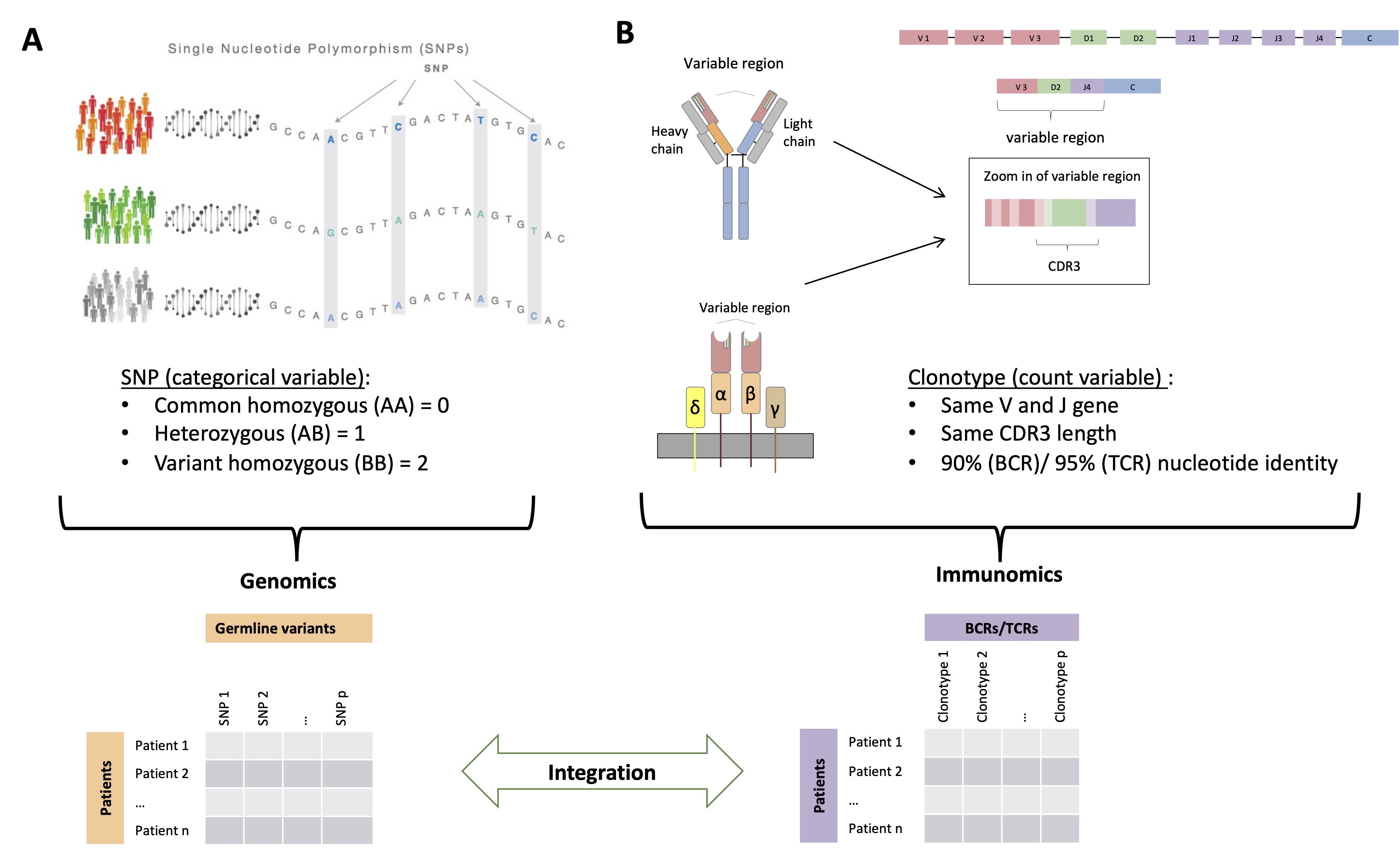}
\caption{Definition and Matrix Data Types. (A) Genomics and (B) Immunomics. }
\label{fig:1}       % Give a unique label
\end{figure}

The adaptive immune system is composed of B and T lymphocytes which produce B cell receptors (BCR) or antibodies capable of recognizing foreign substances, such as pathogens or viruses, and T cell receptors (TCR) which recognize fragments of antigens presented on the surface of the cells. BCR (most commonly called immunoglobulins, IG) consist of two identical heavy chains (IGH) and two light-chains, Kappa (IGK) and Lambda (IGL). Human T-cell receptors (TCR) consist of an alpha and beta chain (TRA and TRB) and a gamma and delta chain (TRG and TRD). The intact antibody contains a variable and a constant domain (C). Antigen binding occurs in the variable domain, which is generated by recombining a set of variable (V), diversity (D) and joining (J) gene segments forming the B- and T- cell immune repertoire (IR), and its diversity is mainly
concentrated in the complementary-determining region 3 (CDR3), from now on, this
combination will be defined as V(D)J (see Figure \ref{fig:1}B).

Recent technological advances have made it possible to apply high-throughput sequencing to characterize BCR and TCR \cite{heather2018high-throughput}. In addition, bioinformatic tools for extracting BCR and TCR from bulk RNA-seq have also been developed \cite{bolotin2015mixcr:}. So, BCR and TCR sequencing allows for a broad examination of B cells and T cells by following changes in clonal and population dynamics. Each sequencing read from BCR or TCR can be grouped into clonotypes which is defined by the number of reads having the same V and J segments, same CDR3 length, and 90\%-95\% nucleotide identity between CDR3s. Consequently a matrix by samples and clonotypes is built for further examination. We have shown before comprehensive analysis using ML techniques applied to this data type in cancer \cite{pineda2021tumor-infiltrating,yu2022pan-cancer} and organ transplantation \cite{pineda2019characterizing}, but numerous other examples have been done in multiple sclerosis \cite{palanichamy2014immunoglobulin} or influenza vaccine responses \cite{strauli2016statistical}. There has been some attempts to integrate genomic profiles and immune repertoire data, such as in this example  \cite{iglesia2016genomic} across 11 tumor types showing that high expression of T and B cell signatures predict overall survival. However, the development of new statistical strategies to analyze and integrate this type of data with other omics are still in progress.

\section{Challenges in the omics data analysis}
\label{sec:3}

The ultimate goal of this research is to investigate whether the variability previously observed in the BCR in pancreatic cancer \cite{pineda2021tumor-infiltrating} is associated with germline genetic susceptibility (SNPs).

Genomic studies are complex and the data is large and very heterogeneous, therefore classical statistical assumptions are limited as has been extensively review previously \cite{ritchie2015methods, kristensen2014principles}.
%{\color{red} Añadir estas ref https://www.nature.com/articles/nrg3868, https://www.nature.com/articles/nrc3721}. 
The most classical statistical approach to assess the relationship between genetic variants (SNPs) with the BCR measuring tumor immune-infiltration is a linear regression model, but one of the main assumptions for this model is the independence between the regressor variables. SNPs are variables that can be highly correlated due to linkage disequilibrium which is the non-random association of alleles at different loci in a given population. Moreover, the high dimensionality is also a current problem affecting method convergence and being computation time-consuming. 

To deal with these problems, ML algorithms are proposed. For example, regularized regression methods (such as the LASSO Regression) are employed when there are many features (more than samples in the study, i.e., $p>>n$, where $p$ and $n$ refer to the number of variables and instances, respectively) and a multicollinearity problem among them. Approaches such as Random Forest consider possible interactions among the features and have increased in popularity in the last few years \cite{li2022benchmark}. 

In complex models, it is very important to propose as parsimonious models as possible. Thus, in ML techniques, feature selection remains a big challenge to be addressed. Additionally, in omics analysis there is a special interest in finding the most influential variables to furtherly extract the biological mechanism that might be involved with the disease under investigation. Currently there are some lines of investigation \cite{cai2018feature} but given the special nature of genetic data, new proposals are needed. Another important challenge relates to the nature of the clinical endpoints frequently used in biomedical studies. That is, the majority of the studies considered dichotomous clinical variables measuring whether the individuals have or have not the disease, or any other specific characteristic, but probably subtypes of clinical outcomes should be contemplated. Additionally, statistical modeling is also designed for dichotomous dependent variables, but the majority of the time, the modeling with clinical outcomes is more complex than just two categories and the implication of ML techniques capable of working with multinomial models are needed.

Taking into consideration these main challenges, in this chapter, we review two of the most commonly used ML techniques and we propose the application of association rules (AR) with predictive purposes extracting the most influential variables, as it will be later shown. Although AR has been previously applied to GWAS studies \cite{nguyen2018detection,creighton2003mining,qian2022association}, as far as we know, they have not been implemented in the predictive context. In addition, we will focus on the application of predictive methods to variables with more than two outcomes, which are less common in the applied literature.

\section{Machine learning techniques}
\label{sec:4}

The term Machine Learning (ML) refers to a broad range of statistical and computational techniques that aim at extracting knowledge from large amounts of data. For that reason, ML methods are specially useful for multivariate approaches, which are needed for the integrative analysis of omics data types. ML techniques can be used for supervised or unsupervised learning; examples of the first category are association and prediction analysis, whereas clustering and dimensionality reduction approaches belong to the latter.

In this chapter we restrict ourselves to the supervised context, where the objective is to develop a tool that allows \textit{guessing} the value of a variable (usually referred to as target) by means of a set of other variables (usually referred to as input). As it will be shown in the following subsections, ML techniques are suitable for this task, especially when the number of input variables is large, the number of instances in the data set is extreme or the input variables are correlated, among others. More precisely, we focus on the methods that can be applied when the target variable is a categorical one with $K >2$ levels (usually referred to as multinomial).

Many different ML methods have been previously proposed in the literature (see \cite{hastie2009elements} for a review), however, none of them has been shown to be capable of coping with the vast amount of data types and applications that arise in the real world. Taking into account the characteristics of omics data, we now review three of the most convenient methods that can be applied in this context: Random Forest, Multinomial Logistic Regression and Association Rules.

\subsection{Random Forests}
\label{subsec:4.1}

Random Forests (RF) were first proposed in \cite{breiman2001random} as an attempt to reduce the large variability associated with classification and regression trees. Trees have been shown to be robust to the presence of missing and outlier values, do not require previous variable selection, work efficiently with both quantitative and qualitative input variables and are very intuitive, making them a very attractive predictive tool. Nonetheless, trees suffer from large variability (as a result of its building process) and are very sensitive to the presence of overpowering input variables (see \cite{hastie2009elements}).

RFs provide predictions by means of a (large) set of trees that have been built from bootstrap samples\footnote{Bootstrap samples are simply random samples drawn with replacement from the original data set that allow emulating the process of obtaining new data samples.}, where the set of input variables used to grow the branches of the trees is randomly restricted. Once the trees are built, predictions are obtained as the average of the predictions generated by each tree.

\begin{figure}[b]
\centering
\includegraphics[scale=.3]{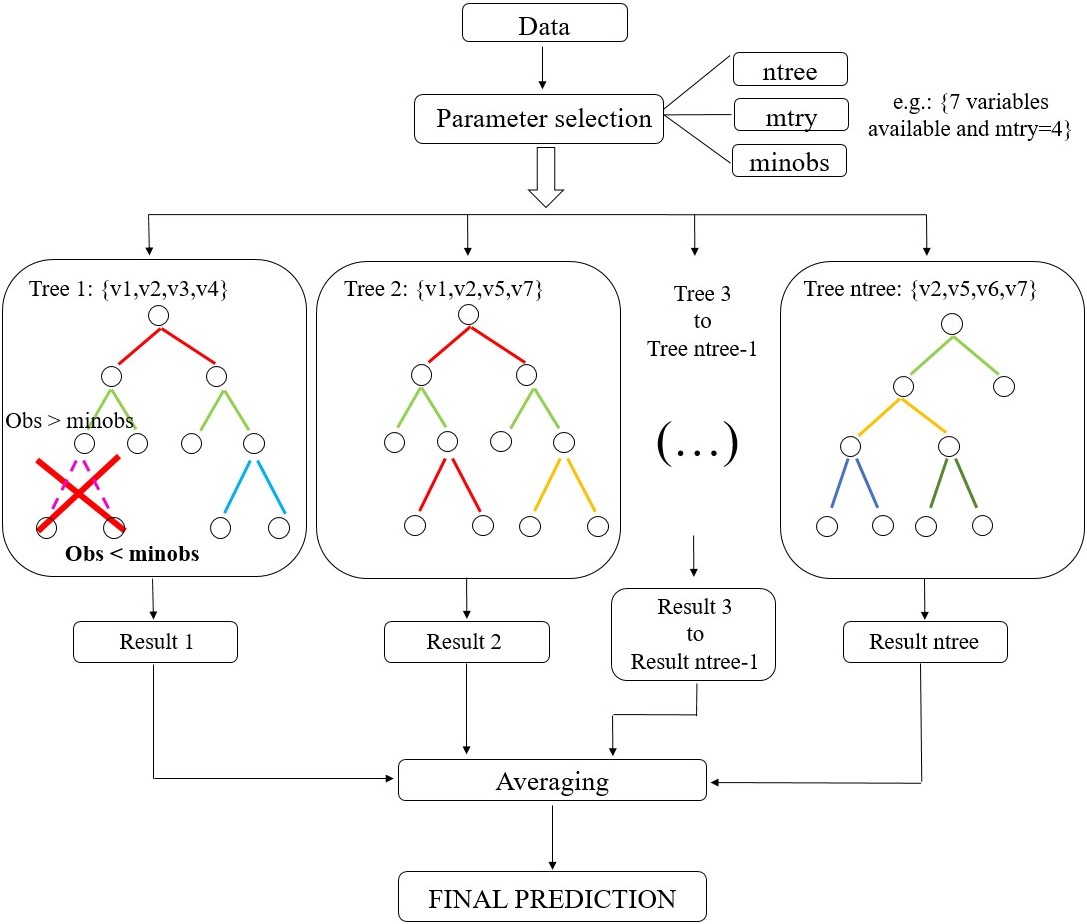}
\caption{Random Forest illustration.}
\label{fig:2}       % Give a unique label
\end{figure}

Including two sources of randomness in the growing process leads to two important characteristics of RF (in comparison with single trees): 1) the variability of the model is reduced because of the properties of the sample mean and the \textit{de-correlation} of the forest attained; and 2) most of the variables with some predictive power make part of the forest, leading to better model performance. Figure \ref{fig:2} shows an schematic illustration of the building process of a RF.

% \begin{figure}[b]
% \includegraphics[scale=.34]{Images/RF1.jpg}
% \caption{Random Forest illustration.}
% \label{fig:2}       % Give a unique label
% \end{figure}

Moreover, RFs have been shown to cope effectively with large data sets, mainly because the trees are built independently of each other and, thus, parallelization can be applied, reducing significantly computational times. Furthermore, no requirement exists regarding the fact that the number of variables is higher than the number of instances available.

Another important feature of RFs, that comes from the properties of the trees in the forest, is the fact that they take into account interactions among input variables. Interactions arise when the input variables are not independent and the effect that they have on the target variable depends on the values that other input variables have. Interactions are made of two or more variables and are usually difficult to find with other types of models, as will be seen in the next section, especially if more than two input variables are involved.

When building RFs, three parameters need to be selected:

\begin{itemize}
    \item The number of trees that make part of the forest (\textit{ntree}). Although RFs do not depend heavily on this parameter, as long as it is high enough, building too many trees can lead to unnecessary large computation time.
    \item The size of the input variable set from which the branches of the trees are built (\textit{mtry}). This parameter has a great impact on performance and computation time, as it affects the way input variables are incorporated in the trees.
    \item The minimum number of observations that a leaf needs (\textit{minobs}). If this parameter is set too small/large, the trees may be very large/small leading to worse performance of the forest.
\end{itemize}

As in any other ML technique, in order to determine the best parameters, in the sense that a good performance is achieved without risking overfitting, resampling techniques are highly recommended (see \cite{hastie2009elements} for more information on this topic). These techniques allow evaluating models with a set of data that did not belong to the training set and, thus, can provide more accurate and reliable estimates of the predictive power of the models. In the context of RFs, a very useful tool is the Out Of Bag (OOB) observations, which allow evaluating the model accurately while building it. As already mentioned, RFs make use of bootstrap samples in the building process and, therefore, every time a tree is grown, a subset of observations are left aside (these are the OOB ones). These data points can be used to assess performance if they are predicted using only the trees in which they did not make part of the growing process. OOB observations are one of the main advantages of RFs as they permit speeding up the process of setting the best parameters.

\subsection{Multinomial Logistic Regression}
\label{subsec:4.2}

Regression models are a wide type of predictive models given by the following equation:
\begin{equation}\label{eq1}
    Y=f(X_1,X_2,\dots,X_p)+\epsilon,
\end{equation}
where $Y$ is the target variable, $\{X_i\}_{i=1,\dots,p}$  are the input variables, $f(\cdot)$ is a mathematical function that relates both types of variables and is usually referred to as link function, and $\epsilon$ is the error term.

Although trees and RFs could be considered a type of regression model, this term is frequently used for the models that can be \textit{easily} written by means of the most suitable link function.

In the special case of the multinomial logistic regression (MLR), where the number of different values of the qualitative target variables equals $K$, with some additional assumptions not included here for the sake of brevity, equation (\ref{eq1}) leads to the following formulation of the model:
\begin{equation}\label{eq2}
    P(y_i=\ell)=\frac{\exp{(\beta_{0,\ell}+\beta_{1,\ell}x_{1,i}+\dots+\beta_{p,\ell}x_{p,i})}}{\sum_{j=1}^K\exp{(\beta_{0,j}+\beta_{1,j}x_{1,i}+\dots+\beta_{p,j}x_{p,i})}},
\end{equation}

\noindent where $i=1,\dots,n$ and $\ell = 1, \dots, K$ and the $\beta$ parameters contain the information of the relationship between the input variables and the target one (see \cite{hastie2009elements} for more details regarding this formulation). Determining the optimal values of these parameters (including which ones are set to 0) is the key of the MLR model.

MLR models have been widely used as they are a natural extension of the classical linear regression model and, thus, their $\beta$ parameters can be interpreted (as opposed to black box models where the exact effect of the input variables on the target remains unknown). However, MLR models do not cope well with outliers and missing observation, which may not be an inconvenience if a proper cleaning of the data is applied. 

Moreover, MLR models do not take into account interactions among input variables unless they are explicitly included in the model as products of them. Furthermore, MLR models assume that the relationship between an input variable and the target one can be modeled through (\ref{eq2}) and, thus, if this assumption does not hold, the predictive power of the model might be reduced.

Finally, and more importantly in this context, the number of parameters to be estimated needs to be strictly smaller than the number of instances available in the training data set. Whereas in the usual application of MLR, this drawback would have little effect (unless interactions are considered), in the context of omics data, most of the application cases deal with data sets with more variables than observations.

In order to deal with this problem and to be able to efficiently select the best input variables (i.e., the ones with corresponding parameters different from zero), some modifications to the classical MLR have been proposed in the literature. In particular, the lasso regularization, and the subsequent group-lasso and sparse-group-lasso have been shown to be very useful.

In the classical MLR context, the optimal parameters are given by the maximization of the likelihood of the training data (or, equivalently, the minimization of the negative-log-likelihood). In this scenario, all the input variables considered get estimated parameters strictly different from zero and, thus, no variable selection is achieved. For that reason, if the number of observations is not sufficiently higher than the number of input variables (as the number of parameters per variable may be higher than one), no solution is found.

The lasso model (acronym for Least Absolute Shrinkage and Selection Operator) was proposed in \cite{tibshirani1996regression} as a solution to the previously mentioned drawback. The basic idea is to add a penalty term in the minimization process that essentially accounts for the number of parameters distinct from zero. The weight given to this penalty term ($\lambda$) allows controlling the number of input variables selected, leading to the classical regression model when it is set to zero.
% It is interesting to note that this penalty term is mathematically equivalent to adding a constraint limiting the number of parameters distinct from zero to the classical optimization problem.
Therefore, the classical lasso model can be stated through the following equation:
\begin{equation}\label{eq3}
    \hat{\boldsymbol\beta}=\underset{\boldsymbol\beta\in \mathbb{R}^{p\times K}}{argmin}\left\{ -\log\mathcal{L}+\lambda\sum_{i=1}^p\sum_{j=1}^K\left|\beta_{i,j}\right|\right\},
\end{equation}

\noindent where $\mathcal{L}$ represents the corresponding likelihood, not shown here for the sake of simplicity, and $\boldsymbol\beta$ is a $p\times K$ matrix containing the $\beta$ parameters.

The original lasso model, although very effective when dealing with data sets where $p>>>n$, is not always the best choice as it can lead to models where only certain parameters of an input variable are selected (and not all of them). In order to avoid this, the group-lasso was proposed \cite{meier2008group} where a modification of the original penalty term is suggested that forces the minimization process to select all or none of the parameters in a group. Generally, the groups are defined by means of the input variables the parameters are associated to, but it is not the only case.

The group-lasso might be too restrictive in some cases leading to models with less predictive power. For that reason, a third alternative was proposed, referred to as sparse-group-lasso \cite{vincent2014sparse} , where both penalty terms are added to the minimization process, with a weight of $\alpha$ to the lasso and $1-\alpha$ to the group-lasso, with  $\alpha\in[0,1]$. For that reason, the sparse-group-lasso can be seen as a generalization to the previous methods, leading to the classical lasso when $\alpha=1$ and to the group-lasso when $\alpha=0$.

In order to determine the best MLR model (with the corresponding lasso regularization) the best combination of $\alpha$ and $\lambda$ needs to be established. As opposed to the RF, in the MLR context there are no OOB observations and, thus, a different resampling technique needs to be applied. In the context of ML techniques, it is very common to resort to $k$-fold cross-validation ($k$-CV), where the data set is divided in $k$ parts and, successively, one is devoted to evaluate the model and the remaining $k$-1 are used to train the model. When the process has been repeated $k$ times, all of the observations in the data set have been predicted once (without having been part of the training process) and that information can be used to accurately evaluate the model.

As a final remark to LASSO-based models, we highlight that this type of models are limited to the sample size as the maximum number of parameters that can be set different from zero.

% If the previous process is repeated with different random partitions, the results become more reliable and lead to what is called repeated cross validation (RCV).

\subsection{Association rules}
\label{subsec:4.3}

Although association rules (AR) are not usually considered supervised models, they can be used to predict the value of a target variable using the values of the input ones, leading to what is usually referred to as Rule-Based Classifiers \cite{liu2001classification,lawrence2001rule-based}. It should be noted that this approach is not very frequently used in the GWAS literature and is, therefore, one of the main contributions of this chapter.

AR are commonly applied in the retail context and, for that reason, its generation is usually referred to as \textit{Market Basket Analysis (MBA)}, as they were first proposed for determining the products in a market basket that were frequently bought together \cite{agrawal1993mining} .

Applying the AR terminology, AR mining consists of detecting sets of items (which can be levels of a variable) that have a ``large" joint frequency and evaluating the co-occurrences found afterwards. More precisely, these co-occurrences are given by ``if-then" statements, where the if clause is denoted as ``lhs" (left hand side) and the consequence, as ``rhs" (right hand side). In the MBA context, an example of rule would be ``If milk and eggs are bought, flour will be as well”. Thus, translating into genetic studies, the rule could be, if a variation on SNP rs1 and rs3 is present, SNP rs5 will have one as well\footnote{Please, note that the usual way of referring to SNPs in the literature is by means of  ``rs" followed by its corresponding identification in numbers.}. 

The aim of AR is, thus, finding rules that take place frequently and can be considered reliable. In order to evaluate AR, three new concepts arise:

\begin{itemize}
    \item Support: the support of a rule is simply the joint relative frequency of the items that belong to the rule. It can be interpreted as the joint probability of the items in the rule. The support is a measure of how often a rule can be used.
    \item Confidence: the confidence of a rule is the probability of the consequence (rhs) conditional to the antecedent (lhs). The confidence of a rule evaluates how often it is true.
    \item Lift: the lift is the ratio of confidence to support. This measure compares how often the lhs and the rhs occur together in comparison with what should be expected if both were independent. Alternatively, it quantifies how often the rhs arises along with the lhs in comparison with the ``general population".
\end{itemize}

Ideally, all combinations of items should be generated and evaluated in order to generate a list of useful ones. However, as the full set of items is generally very large, this procedure is usually unfeasible. For that reason, several procedures have been proposed in the literature to find AR in a data set (see \cite{KotsiantisS2006} for a review on association rules mining). In this chapter we restrict ourselves to two of them: the \textit{apriori} algorithm and the Random Forest.

The \textit{apriori} algorithm (proposed in \cite{agrawal1994fast}) assumes that, in order to efficiently generate the rules, the ones that do not appear frequently enough should be skipped as they will not be useful. For that reason, instead of generating all possible combinations of items, the algorithm sequentially generates larger rules combining previous sets that have a sufficiently high support (called frequent itemsets). Applying this simple idea, computation time are drastically reduced without risking removing useful rules (as frequent itemsets are always made of frequent “itemsubsets”). Moreover, rules that do not reach a minimum confidence level are removed, reducing the storage requirements.

This algorithm was initially proposed in the context of MBA and, thus, if one wants to apply it to generate AR more generally, the variables involved in the rules to be generated need to be transformed previously. More precisely, each variable needs to be converted to an item in such a way that, instead of working with items purchased together, we deal with values of variables that take place in the same observation. For that reason, qualitative variables need to be converted to dummy variables and quantitative variables need to be first discretized and then converted to dummy variables as well. Regarding quantitative variables, we note that some loss of information takes places because of the discretization process and, thus, this process needs to be done carefully.

Although the \textit{apriori} algorithm significantly reduces computation time, if the number of variables is very large (and, thus, the number of items is even larger), it still can become unfeasible to generate the rules as the process cannot be parallelized. In such cases, one can resort to different methods of variable selection to reduce the number of variables (or equivalently items) involved.

Alternatively to the \textit{apriori} algorithm, the RF method assumes that the leaves of the trees in the forest can be considered as rules \cite{sirikulviriya2011integration,bostrom2018explaining}, as the conditions that define the leaf might serve as antecedents and the majority target classes, as their consequences. Once generated, the rules are analyzed and the ones not reaching a sufficient confidence level are discarded.

This method has several advantages: the first one deals with the fact that input variables do not need to be converted to dummy variables and, thus, the items considered can be made of several levels of the same variable. Finally, as the procedure can be parallelized, it is very fast in comparison with the \textit{apriori} algorithm. However, because of the randomness included on the input variables, not all combinations of them are evaluated and, thus, some powerful rules might never be found. It is important to highlight that this method is only applicable when the only desired rules are the ones that involve a single variable (in our case the target one) as consequence.

As previously mentioned, AR can be used with predictive purposes if the rules obtained contain one of the levels of the target variable as the consequence. In this sense, the observations that fulfill the antecedent condition, can be classified as the consequence. 

Classical predictive models need to provide a procedure that allows classifying all observations in a data set. However, some observations are generally more difficult to predict than the rest and, because of that, if models are evaluated considering all the available observations it might seem that the models are useless or, in other words, that the input variables do not have information on the target one. AR permit focusing only on the ``strong" associations between input and target variables, even if they do not involve all observations in the data set, leading to some insight on the relationship between input and target variables. Moreover, the antecedent can contain several input variables, allowing to include interactions among them.

\section{Application}
\label{sec:5}

\subsection{Study subjects}
\label{subsec:5.1}

The data used in this chapter come from a total of 144 confirmed PC cases from TCGA \cite{peran2018curation} \footnote{\url{https://portal.gdc.cancer.gov/}}.

The BCR data was extracted from the RNA sequencing (RNA-seq) data using MiXCR tool \cite{bolotin2015mixcr:}, which align the RNA-seq data in \textit{fastq} format (storage of sequence data) to the VDJ region to extract IGH, IGK, IGL and TRA, TRB, TRD and TRG. In previous analysis \cite{pineda2021tumor-infiltrating}, we found IGK clonotypes (defined by the same V and J gene, same CDR3 length, and 90\% nucleotide identity) discriminating PC samples. Using a centered log ratio (CLR) LASSO for compositional data followed by a hierarchical clustering, three main clusters were found to classify the PC samples. The distribution of samples across the three clusters was: cluster 1 (22,43\%), cluster 2 (43,93\%) and cluster 3 (33,64\%). Cluster 1 was characterized by a lower infiltration and was more similar to the normal pancreas associated with higher tumor purity and worse survival while cluster 3 was the one with higher infiltration and better survival. Cluster 2 was closer to cluster 3 but showing some differences in tumor purity and survival. We now believe that these clusters might be explained by germline genetic variation as stated in the introduction
\cite{orru2013genetic, patin2018natural, liston2018origins}. Therefore, for the purpose of this chapter, we downloaded the genotyping data available from the same set of patients in TCGA. A total of 906,600 blood genotypes were available from 120 PC samples. The monomorphic SNPs and the ones with minor allele frequency $< 0.05$ were excluded. To avoid high multicollinearity, those SNPs that were in Linkage Disequilibrium (LD) were also excluded. We used the D coefficient to obtain the LD \cite{slatkin2008linkage} and excluded those that were in high LD ($>0.8$). In this data set, patients come from different races, to avoid population stratification, we selected only the Caucassian individuals ($n=107$ individuals). The final data set was composed of a total of 117,486 SNPs and 107 PC samples and SNPs were defined as $0,1,2$ corresponding to the number of variant alleles per sample. 
%({\color{red} ref: https://www.ncbi.nlm.nih.gov/pmc/articles/PMC5124487/)}

\subsection{Material \& Methods}
\label{subsec:5.2}

Taking all previous information into consideration, the aim of this application is to develop a predictive tool considering the set of SNPs that allows to classify PC patients into one of the 3 previously mentioned clusters, and give some insight on the most influential SNPs. For that purpose, we will make use of LASSO MLR models, RF and AR.

As already mentioned, when applying prediction methods, the use of resampling techniques is widely spread as it permits obtaining more accurate estimates of the prediction power. In this chapter, we will partition the data set into training (80\% of the samples) and testing (the remaining 20\%) in order to develop and evaluate the models, respectively.

When dealing with multinomial models, the most frequent measure of evaluation is the accuracy, which is simply the percentage of observations that have been correctly classified. However, in the presence of unbalanced target variables, this measure can lead to misleading conclusions as the best results can be achieved by classifying all observations into the larger category from the target variable. In order to avoid this problem, alternative evaluation measures have been proposed and analyzed \cite{FerriC2009} that focus mainly on the predicted probability of each of the categories. In this chapter, we make use of the pairwise AUC (Area Under the receiver operating characteristic Curve), which is obtained as the average of the AUCs obtained from the predictive probabilities considered in a one vs. one scenario. On the other hand, regarding the AR, we will make use of the three previous measures, i.e., support, confidence and lift.

Regarding the software used for this application, we make use of R 4.2.0 and, more precisely, packages \textit{randomForest}, \textit{msgl} and \textit{arules} for the prediction part. For the performance evaluation, package \textit{pROC} provides the pairwise AUC and we have adapted the code of package \textit{OOBCurve} for the OOB resampling technique. Moreover, the CPU time required to train the models on a HP ProDesK 400 G7, Intel(R) Core(TM) i7-10700 CPU @ 2.90GHz, RAM: 16GB (8 × 2GB), was 19 seconds for the RF (parallelizing the process) and 125 seconds, for the LASSO MLR.

\subsection{Results}
\label{subsec:5.3}

\subsubsection{Random Forest and LASSO Multinomial Logistic Regression}

\begin{figure}[b]
\includegraphics[width=\textwidth]{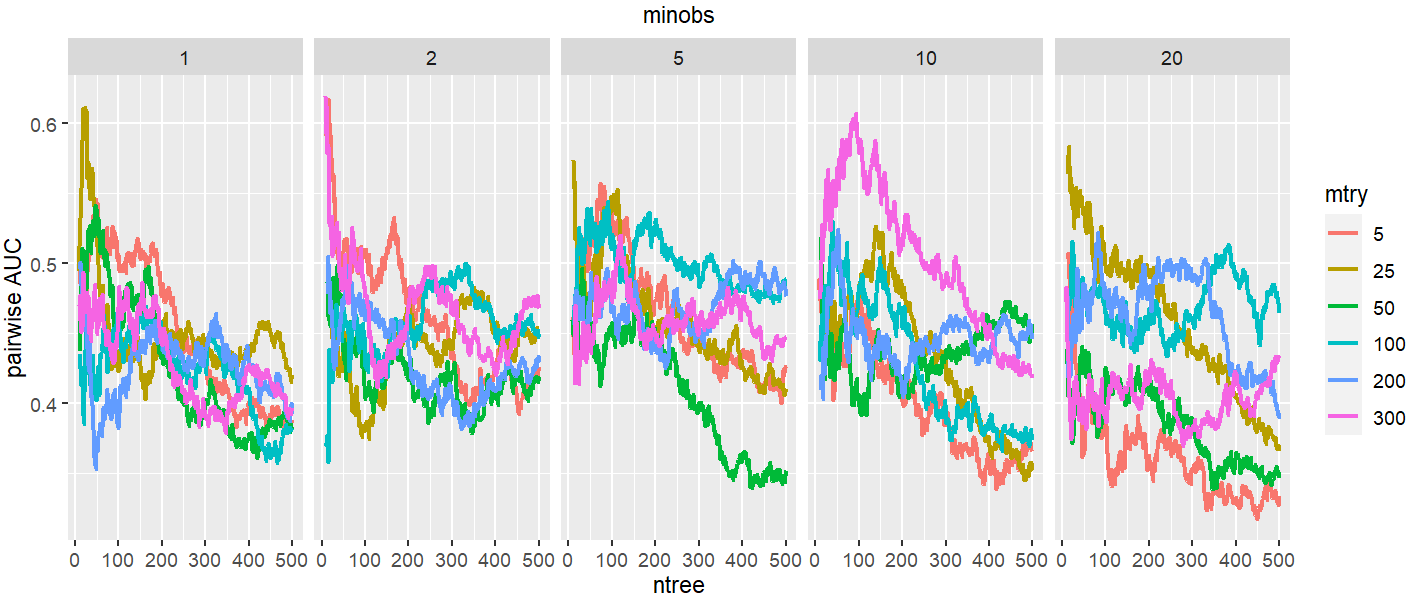}
\caption{Pairwise AUC obtained from the OOB observations for the different configurations of the RF.}
\label{fig:3}       % Give a unique label
\end{figure}

In order to find the most suitable predictive model for a data set, the corresponding parameters need to be tuned. For that purpose, as already mentioned, we resort to the OOB observations in the RF scenario, and the $k$-CV scheme for the LASSO MLR applied only to the train partition of the data.

\begin{figure}[t]
\includegraphics[width=\textwidth]{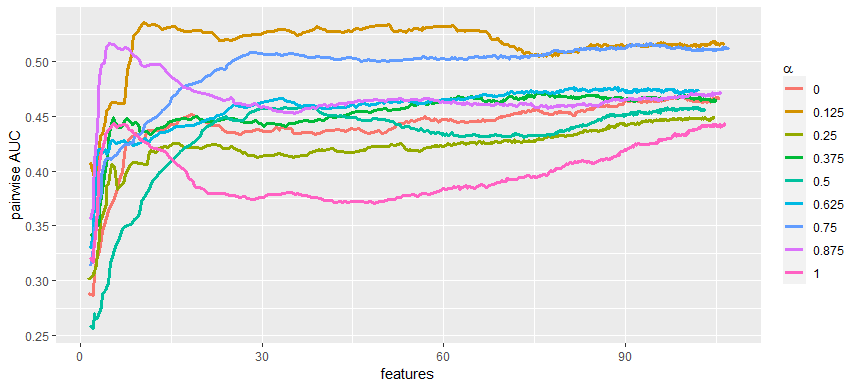}
\caption{Pairwise AUC obtained from $k$-CV for the different configurations of the LASSO MLR.}
\label{fig:4}       % Give a unique label
\end{figure}

Figures \ref{fig:3} and \ref{fig:4} show the pairwise AUC of both types of models for the different parameters that need to be tuned. In particular, Fig. \ref{fig:3} shows the results for the RF parameters, i.e., \textit{minobs}, \textit{mtry} and \textit{ntree}. As it can be seen, there is no clear pattern regarding the best configuration. Moreover, the models do not seem to be very powerful as for many configurations, the AUC is smaller than 0.5. Nonetheless, the best configuration leads to a value slightly larger than 0.6 and corresponds to approximately 100 trees (94 precisely), a minimum of 10 observations per leaf (which corresponds with approximately 10\% of the observations available) and random sets of $300$ variables. Regarding the best setup, Table \ref{tab1} contains information on the performance as well as the number of SNPs (variables) that have been effectively selected by the RF (out of the 117,486 available).

On the other hand, Figure \ref{fig:4} shows the results for the LASSO MLR parameters, i.e., the values of the $\lambda$ and $\alpha$ parameters. It should be noted that, for the sake of understanding, instead of plotting the value of $\lambda$, we have opted for the corresponding number of features selected (that is, the number of variables that have, at least, one parameter different from zero). As it can be seen, this type of model leads to lower predictive power (as the pairwise AUCs obtained take smaller values). Among many other reasons, we believe that the fact that interactions are not taken into account, being frequently expected in the SNPs context, could be one of the main reasons. Nevertheless, we can conclude that the best configuration consists of selecting $63$ features applying a value of $\alpha$ equal to $0.125$. Table \ref{tab1} contains information on the performance of the LASSO (both by means of the $k$-CV and the test partition). 
% The fact that the $\alpha$ parameters takes a value different from zero and one evinces the usefulness of the sparse-group LASSO model in this context. In particular, such a small value means that some flexibility was needed in the group LASSO.

% {\color{red} ¿Comentamos algo más?}

\begin{table}[ht]
\centering
\caption{Performance measures of the best RF and LASSO MLR evaluated in the test partition and the corresponding resampling method. The number of SNPs selected by each model is also shown.}
\renewcommand{\arraystretch}{1.2}
\setlength{\tabcolsep}{5pt}
\label{tab1}
\begin{tabular}{l|c|c|c}
      & \# of SNPs & AUC pairwise (test) & AUC pairwise (OOB/$k$-CV) \\ \hline
RF    & 723    & 0.668               & 0.607                  \\
LASSO & 63     & 0.530              & 0.586                
\end{tabular}
\end{table}
% {\color{red} Me he perdido un poco en los métodos y resultados de AR. Del texto no entiendo muy bien si se aplican los 3 algortimos que luego aparecen en las tablas. Es decir, en el texto aparece RF extraction y a priori algorithm y luego en la tabla aparece también el LASSO MLR a priori, pero en el texto no queda tan claro, entonces por aclarar se  han utilizado 3 métodos: RF extraction, RF a priori (con los SNPs seleccionados por RF en el apartado anterior) y luego LASSO MLR a priori (con los SNPs seleccionado en el apartado anterior con LASSO MLR) es así? por aclararme yo y aclararlo en el texto}

\subsubsection{Association rules}
% {\color{red} he modificado un poco todo este párrafo para que se entienda mejor por si queréis revisarlo pero primero aclararía lo que os pongo más arriba}
As the performance of the previous models is not sufficiently good to apply them in a predictive context, we now generate association rules that can lead to extract new information of the relationship between SNPs and our target variable.
In order to generate the AR, we have applied the two different algorithms previously mentioned (RF extraction and the \textit{apriori} algorithm). For the latter, we have obtained two different sets of rules by working only with the variables selected in the corresponding predictive model (LASSO MLR and RF). The rationale behind this decision is computational as the number of available items (without subsetting) equals to 352,461 (three dummy variables per SNP plus other three for the target variable) and the corresponding number of combinations gets very large.
In this sense, it is important to highlight that the library we have used, \textit{arules}, establishes a maximum time for obtaining the combinations. For that reason, when the total number of items is large, the rules obtained are usually made by only one or two items. Moreover, this function does not allow generating only the rules that have the target variable as a consequence and, for that reason, the computation capability is not maximized. For instance, the RF-\textit{apriori} leads to over 60 million rules, most of which are discarded.

Table \ref{tab2} contains information on the three sets of rules than have been found: two using solely the SNPs selected by the best RF and LASSO MLR separately\footnote{We note that we have not considered the SNPs selected by both methods at the same time because of the time limit previously mentioned, as the number of rules generated was smaller than the sum of the rules obtained separately.} and the one obtained directly from the leaves in the RF. It is important to highlight that we have set a minimum support of $10\%$ (which corresponds to the optimum number of observations per leaf in the RF) and a minimum confidence of $0.7$.

\begin{table}[ht]
\centering
\caption{Summary of the three sets of association rules obtained.}
\renewcommand{\arraystretch}{1.2}
\setlength{\tabcolsep}{3.4pt}
\label{tab2}
\begin{tabular}{c|c|c|c|c|c|c|c}
Method & \begin{tabular}[c]{@{}c@{}}\# of \\ rules\end{tabular} & \begin{tabular}[c]{@{}c@{}}\# SNPs \\ in a rule\end{tabular} & \begin{tabular}[c]{@{}c@{}}Average\\ lift\end{tabular} & \begin{tabular}[c]{@{}c@{}}Max\\ lift\end{tabular} & \begin{tabular}[c]{@{}c@{}}\% of rules\\ target=1\end{tabular} & \begin{tabular}[c]{@{}c@{}}\% of rules\\ target=2\end{tabular} & \begin{tabular}[c]{@{}c@{}}\% of rules\\ target=3\end{tabular} \\ \hline
RF extraction & 115 & 2-8 & 2.014 & 4.458 & 4.35\% & 66.96\% & 28.69\% \\
RF \textit{apriori} & 14482 & 1-2 & 1.740 & 4.087 & 0.19\% & 94.49\% & 5.32\% \\
LASSO MLR \textit{apriori} & 29773 & 1-7 & 2.012 & 4.458 & 0.72\% & 80.38\% & 18.90\%
\end{tabular}
\end{table}

As it can be seen, the number of rules extracted from the RF is significantly smaller than the \textit{apriori} cases; this is due to the fact that not all combinations of SNPs are evaluated and, moreover, the number of selected trees in the forest is set relatively low (94). Despite of that, the strength of the rules is as high as for the rules generated with SNPs selected by the LASSO MLR (both methods lead to similar values of the lift). 

Because of the large number of SNPs selected by the RF, the \textit{apriori} algorithm is not capable of generating combinations involving more than $3$ items. This fact has two main consequences: 1) less complex interactions among SNPs can be found and 2) the rules found are slightly less strong than in the other methods (although still useful).

From the set of SNPs selected by RF and LASSO in the predictive models, less than 1\% are overlapping between the two methods. Therefore, the number of different rules achieved by the three methods altogether reaches 45,000. Importantly, all these rules might be used with predictive purposes and, moreover, can give new biological insight considering the combinations of SNPs that frequently lead to the different levels of the target variables.

Furthermore, Table \ref{tab2} includes information on the percentage of rules having as consequence the three levels of the target variable. The results show that significantly more rules are generated for the second level, which is related to the fact that this level is the most frequent one. Nevertheless, as many rules are generated, even if the percentage is low, the absolute number per level is not negligible. 

Finally, in order to illustrate the rules generated, Table \ref{tab3} shows, considering each of the outcomes of the target variable as consequence (RHS), the rules with the largest lift value for the three methods. As it can be seen, the rules associated with the first level are the "strongest" ones, as the maximum lift value exceeds 4, whereas this maximum value is close to 3 and slightly larger than 2 for the third and second levels of the target variable, respectively. This is in contrast with the number of rules generated, showing that more rules do not necessarily mean better predictive power.

% However, regarding support and confidence, no differences are detected.

% \usepackage{multirow}
\definecolor{mycolor}{gray}{0.85}
\begin{table}[ht]
\centering
\caption{Analysis of the most important rules generated by method and target value.}
\renewcommand{\arraystretch}{1.2}
\setlength{\tabcolsep}{5pt}
\label{tab3}
\begin{tabular}{c|l|c|c|c|c}
Method & Antecedent SNPs & Target value & Support & Confidence & Lift \\ \hline
 & \cellcolor{mycolor} \begin{tabular}[c]{@{}l@{}}rs8313=0, rs38334\textgreater{}=1,\\ rs602560\textgreater{}=1, rs81162=0,\\ rs97842\textless{}=1\end{tabular} & \cellcolor{mycolor}1 & \cellcolor{mycolor}0.112 & \cellcolor{mycolor}1 & \cellcolor{mycolor}4.458 \\
\multirow{2}{*}{RF extraction} & \begin{tabular}[c]{@{}l@{}}rs1707\textgreater{}=1, rs43904\textgreater{}=1, \\ rs45962\textgreater{}=1\end{tabular} & 2 & 0.131 & 0.933 & 2.125 \\
 & \cellcolor{mycolor}\begin{tabular}[c]{@{}l@{}}rs12902\textless{}=1, rs38757\textless{}=1, \\ rs49632\textless{}=1, rs81515\textless{}=1, \\ rs92454\textgreater{}=1, rs107661=0, \\ rs113085\textgreater{}=1\end{tabular} & \cellcolor{mycolor}3 & \cellcolor{mycolor}0.168 & \cellcolor{mycolor}0.947 & \cellcolor{mycolor}2.816 \\ \hline
\multirow{3}{*}{RF \textit{apriori}} & rs8298918=0, rs8633827=2 & 1 & 0.103 & 0.917 & 4.087 \\
 & \cellcolor{mycolor}rs8519341=1, rs8541272=1 & \cellcolor{mycolor}2 & \cellcolor{mycolor}0.103 & \cellcolor{mycolor}1 & \cellcolor{mycolor}2.277 \\
 & rs8536395=0, rs8717271=0 & 3 & 0.121 & 1 & 2.972 \\ \hline
\multirow{4}{*}{\begin{tabular}[c]{@{}c@{}}LASSO\\ MLR\\ \textit{apriori}\end{tabular}} & \cellcolor{mycolor}\begin{tabular}[c]{@{}l@{}}rs4230313=0, rs8348471=1, \\ rs8559618=0, rs8674253=0\end{tabular} & \cellcolor{mycolor}1 & \cellcolor{mycolor}0.103 & \cellcolor{mycolor}1 & \cellcolor{mycolor}4.458 \\
 & rs1858797=2, rs8629623=1 & 2 & 0.103 & 1 & 2.277 \\
 & \cellcolor{mycolor}\begin{tabular}[c]{@{}l@{}}rs8634851=0, rs8661371=0, \\ rs8700294=0\end{tabular} & \cellcolor{mycolor}3 & \cellcolor{mycolor}0.121 & \cellcolor{mycolor}1 & \cellcolor{mycolor}2.972
\end{tabular}
\end{table}

As an example of the rules obtained, we now comment in detail two of the rules in Table \ref{tab3}:
\begin{itemize}
    \item The first rule included in the table for the \textit{RF apriori} method states that the patients that have the two common alleles for the SNP rs8298918 and the two variant alleles for the SNP rs8633827, will belong to the first level of the target variable. This combination of facts happens in 10.3\% of the patients and holds in 91.7\% of the cases. Regarding the interpretation of the lift, patients that have the two common alleles for the SNP rs8298918 and the two variant alleles for the SNP rs8633827 are 4.09 times more likely to belong to the first level of the target variables than a randomly selected patient. A similar analysis can be carried out for the remaining rules obtained through the \textit{apriori} algorithm).
    \item The second rule shown in the table for the RF extraction procedure can also be analyzed using the previous ideas regarding the values of support, confidence and lift. However, it should be noted that, as previously mentioned, rules obtained directly from trees have the advantage of being more complex as several values of a SNP can be considered. In particular, this rule states that patients with at least one variant allele for the SNPs  rs1707, rs43904 and rs45962 will belong to the second level of the target variable.
\end{itemize}

\section{Conclusions and future work}
\label{sec:6}
In this work, we have made a review on the use of ML techniques to integrate omics data with a special focus in predictive modelling. Moreover, we have illustrated the main ideas using a real data set composed by 107,486 blood genotypes as predictor variables and a 3-categorical variable measuring the immunological infiltration in pancreatic tumors as target variable in a total of 107 cleaned samples. 

First, we have showed the performance of the RF and LASSO MLR algorithms being not sufficiently good. They showed maximum AUCs of 0.6 in their best parameter configuration. Interestingly, for this part, is the fact that among the $723$ SNPs selected by the RF and the $63$ selected by the LASSO MLR, only 6 of them overlap, which exposes the differences between both types of models. LASSO MLR models are more rigid and, thus, tend to select those SNPs that better adjust to the assumed formula (see Eq. (\ref{eq2})), whereas RF are more flexible and take into account the interactions among them. Thus, the highest AUC observed is using the RF algorithm probably due to this fact.

In omics studies, especially when using genomic data, the predictive ability normally is not enough to obtain good predictive models, therefore we sometimes need to restrict ourselves to association models looking for the most important variables contributing to the target variable. Here, we propose the use of AR to predict the value of the target variable using the blood genotypes, obtaining important rules with good numbers of support, confidence and lift. Furthermore, we have used different algorithms to extract the most significant rules showing good performances with all of them. 

Our results are sound and revealing and propose a new way to make predictions using AR that extracts the most relevant information between the SNPs and the target variable in a more flexible predictive context. As future work, we are planning to apply the rules obtained here in a complete independent data set to evaluate the predictive accuracy of this new tool. Moreover, we plan to investigate the biological mechanism behind these SNPs selected for each cluster.

\begin{acknowledgement}
We thank Núria Malats, Marina Sirota and Laura Gutierrez for their involvement with the data generation. We also thank to the grant supported this work (the 2019 AACR-AstraZeneca Immuno-oncology Research Fellowship (19-40-12-PINE)).
\end{acknowledgement}

{\small\rm
\noindent\textbf{Author Contribution Statement} 
SP and AC conceived the study design and analysis plan. SP extracted and curated the data. AC performed the data analysis. AM designed the tables and figures. SP supervised the work. All authors
contributed to the article written and approved the submitted version.}

% \newpage

\section*{Appendix}
\addcontentsline{toc}{section}{Appendix}

\begin{table}[h]
\centering
\caption{Acronyms and their respective meaning.}
\renewcommand{\arraystretch}{1.2}
\setlength{\tabcolsep}{8pt}
\label{tab4}
\begin{tabular}{l|l}
\multicolumn{1}{c|}{Acronym} & \multicolumn{1}{c}{Meaning} \\ \hline
AR & Association Rules \\
AUC & Area Under the receiver operating characteristic Curve \\
BCR & B Cell Receptors \\
CDR3 & Complementary-Determining Region 3 \\
CLR & Centered Log Ratio \\
% D & Diversity \\
DNA & Deoxyribonucleic acid \\
eQTL & expression of Quantitative Trait Loci \\
GWAS & Genome-Wide Association Studies \\
IG & Immunoglobulins \\
IGH & Immunoglobulins with Heavy chains \\
IGK & Immunoglobulins Kappa \\
IGL & Immunoglobulins Lambda \\
IR & Inmune Repertoire \\
% J & Joining \\
$k$-CV & $k$-fold Cross-Validation \\
LASSO & Least Absolute Shrinkage and Selection Operator \\
LD & Linkage Disequilibrium \\
MBA & Market Basket Analysis \\
ML & Machine Learning \\
MLR & Multinomial Logistic Regression \\
OOB & Out Of Bag \\
PC & Pancreatic Cancer \\
RF & Random Forests \\
RNA & Ribonucleic Acid \\
SNPs & Single Nucleotide Polymorphisms \\
TCGA & The Cancer Genome Atlas \\
TCR & T Cell Receptors \\
TRA & T cell Receptors Alpha \\
TRB & T cell Receptors Betha \\
TRD & T cell Receptors Delta \\
TRG & T cell Receptors Gamma \\
% V & Variable \\
V(D)J & Variable (Diversity) Joining combination
\end{tabular}
\end{table}

% %
% %
% When placed at the end of a chapter or contribution (as opposed to at the end of the book), the numbering of tables, figures, and equations in the appendix section continues on from that in the main text. Hence please \textit{do not} use the \verb|appendix| command when writing an appendix at the end of your chapter or contribution. If there is only one the appendix is designated ``Appendix'', or ``Appendix 1'', or ``Appendix 2'', etc. if there is more than one.

% \begin{equation}
% a \times b = c
% \end{equation}

\bibliographystyle{ieeetr}
\bibliography{bibliography}

\end{document}